\documentclass[%
 reprint,
 amsmath,amssymb,
 aps,
]{revtex4-2}

\usepackage{graphicx}
\usepackage{dcolumn}
\usepackage{bm}
\usepackage{hyperref}
\usepackage{subcaption}
\usepackage{placeins}
\usepackage{anyfontsize}

\usepackage[font={footnotesize,it}]{caption}

\begin{document}

\preprint{APS/123-QED}

\title{Agent-Based Model of Crowd Dynamics in Emergency Situations:\\ A Focus on People With Disabilities }

\author{Janey Alex}
 \email{jane.alex@uvm.edu}
 
\author{Jason Stillerman}%
 \email{jtstille@uvm.edu}

\author{Noah Fritzhand}%
 \email{nfritzhand@uvm.edu}
\author{Tucker Paron}
 \email{tparon@uvm.edu}

\date{\today}

\begin{abstract}
A link to all of the code: \url{https://github.com/Stillerman/crowd-dynamics.git}

\end{abstract}

\maketitle

\section{Overview and Motivations}
Collective behavior of people in large groups and emergent crowd dynamics can have dangerous and disastrous results when panic is introduced. These events can be caused by emergency situations such as fires in a large building or a stampeding effect when people are rushing in a densely packed area. In this paper, we will use an agent-based modeling approach to simulate different evacuation events in an attempt to understand what is the most efficient scenario. Specifically, we will focus on how people with disabilities are impacted by chosen parameters during an emergency evacuation. We chose an ABM to simulate this because we want to specify specific roles for different “agents” in our model. Specifically, we will focus on the influence of people with disabilities on crowd dynamics and the optimal exits. Does the placement of seating for people with disabilities affect the time it takes for the last person to exit the building? What effect does poor signage have on the time it takes for able-bodied and people with disabilities to exit safely? What happens if some people do not know about alternative exits in their panicked state? Using our agent-based model, we will investigate these questions while also adjusting other outside effects such as the density of the crowd, the speed at which people exit, and the location of people at the start of the simulation.

\subsection{Previous Work}
Of our examined papers, ‘Simulating dynamical features of escape panic’ by Dirk Hellbing, Illés Farkas, and Tamás Vicsek [1] proved to be one of the most relevant. The paper introduced the topic of stampeding crowds and identified frequent causes such as panic, eagerness, or randomness. Hellbing, Farkas, and Vicsek aimed to investigate the behavior of crowds in these situations across these different crowd and exit types. As a preemptive analysis, the authors researched related studies across different disciplines and identified 9 critical behavioral findings. Some of these findings include faster and more physical movements as well as a slowing of escape due to injured or fallen people. It also included the unfortunate fact that alternative exits are often overlooked by people in a panicked state. Additionally, authors noted difficulty finding real event data to effectively test their model - something that should certainly be considered and acknowledged for our work as well. Some details of their model that may be useful inspiration are the comparison of individualistic (random) versus herding behavior (and how that pertains to exit success rate) and variance of exit shape and quantity. The authors concluded their study with potential test cases, noting testing building suitability for emergency as a primary potential use. Another important paper that informed our thinking about disabled people in a crowd was 'How people with disabilities influence crowd dynamics of pedestrian movement through bottlenecks' published recently in Nature. In this paper, the results from 12 different studies were analyzed. Each of the 12 detailed findings when people with and without visible disabilities exited through a bottleneck exit. The authors were able to compile a list of three main findings from the papers they reviewed. One, participants with disabilities reduced their speed further away from the bottleneck than participants without disabilities; two, participants without disabilities stayed closer to neighbors with disabilities than to neighbors without disabilities; and three, participants interacted and communicated with each other to organize in front of the bottleneck [2]. These conclusions have helped structure our thinking and assumptions as we began modeling. At the end of our analysis, and throughout our development, it will be important to consider how our model can be of use to others.

\section{Methods}

\subsection{Model Assumptions}
\indent Before fully implementing the agent-based model, there were a number of assumptions that had to be made about both the environment and the nature of the agents themselves. To simplify the modeling process, the environment was modeled as an empty room, a flat rectangular grid without any obstructions. Additionally, there would be no mid-evacuation events and agents have no attachments among one another. Another big assumption built into the model is that all the agents move toward and exit at their closest exit. This means that there is no decision-making architecture for agents to change which exit they are going toward based on their surroundings. \\ \indent The choice to specifically model agents with disabilities came with another set of decisions. Building off previous work and our own experiences, we chose to make all agents, with or without visible disabilities, move at similar speeds ($\pm 5\%$). Furthermore, we chose to give agents with disabilities a larger radius and more mass associated with them. These decisions were made under the assumption that those with physical disabilities may be using assisted walking devices such as a wheelchair, crutches, or a walker. We were working under the assumption that these items take up more space but also people might have a higher tendency to move out of the way when a wheelchair is coming through. Giving them more mass achieves this effect in the physics engine.

\subsection{Model implementation}
\subsubsection{Design of the data structure used to store attributes of the agents} Using physics-based engines Pymunk and Pygame, we created a \texttt{People} class. Each person object contains an x and y coordinate, values to designate shape, mass, and speed. Additionally, each person contains boolean values designating whether they are disabled, if they can use all exits, and if they can fall down during the evacuation. Agents are then stored in a Python list of people objects during the simulation of the crowd. 
\subsubsection{Design of the data structure used to store the states of the environment}
The state of the environment is the location of the exits and the walls. The exits are in the four corners, with two potential additional ones on the side walls. 
\subsubsection{The rules for how the environment behaves on its own}
The environment is static, it doesn't change over time. 
\subsubsection{The rules for how the agents interact with the environment}
Agents are attracted to the exits and cannot go through the walls. 
\subsubsection{The rules for how the agent behaves on their own}
An agent does not interact with itself unless we set the experiment up so that the agent can fall. If there were no exits, no other people, and agents cannot fall, the agent would just stand still. With falling enabled, an agent has a 1/1000 chance each time step of falling and then a 1/100 chance each time step thereafter of getting back up.
\subsubsection{The rules for how agents interact with each other}
These rules are primarily determined by Pymunk. Agents cannot overlap but do not repel one another. 

\section{Results}

\subsection{Control}
\begin{figure}[!htbp]
     \includegraphics[width=6cm]{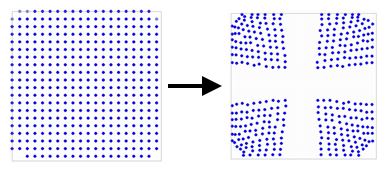}
 \caption{Control layout for crowd evacuation simulation. 400 people, 1 exit in each corner.}
 \label{control}
 \end{figure}

The control model consisted of 400 identical agents initialized in a grid formation with exits in each of the four corners of the "room" (see Figure \ref{control}). A graph showing the number of agents left after any given number of time steps would show a predominantly linear relationship for the control scenario. This is due to the fact that all the agents are moving at the same speed and have the same mass. Because of this, it is effectively as if the agents are not interacting at all. That is until they get to the exit and are possibly bumped by a neighbor. 

\subsection{People With Disabilities and Designated Locations}
\begin{figure}[!htbp]
     \includegraphics[width=8.9cm]{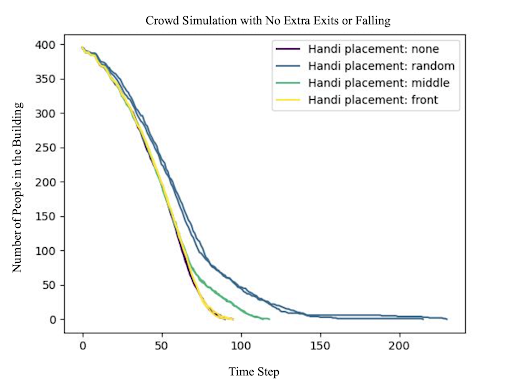}
 \caption{Designated placement of people with disabilities in crowd of 400.}
 \label{standard}
 \end{figure}

\indent When including people with disabilities and designating specific locations for these groups to view, there emerges a most optimal layout. When having a section dedicated to people with disabilities in the front of the 400-person crowd, there is little to no difference in evacuation time from our control group (which does not include people with disabilities). The other placements, middle and random, result in sub-optimal evacuation times which our random placement having a particularly long tail as crowd members without disabilities get stuck behind our group with disabilities due to their larger encoded radius. \\
\indent Each group is simulated five times as is displayed in Figure \ref{standard}. This is the case for all three of the tests and figures.
 
\subsection{Extra Exits}
\begin{figure}[!htbp]
     \includegraphics[width=8.9cm]{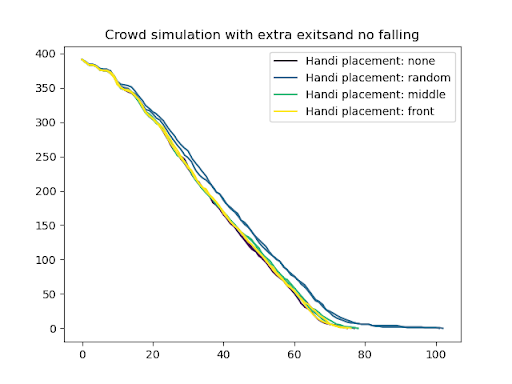}
 \caption{Designated placement of people with disabilities while including extra exits in crowd of 400.}
 \label{extra}
 \end{figure}

\indent Next two extra exits will be added to test the crowd placements under different conditions. With these six exits (as opposed to the standard four), the location of people with disabilities becomes significantly less important. Besides the random placement, middle, front, as well as the group groups, all yield nearly identical results (see Figure \ref{extra}) - each having an average evacuation time of around 80 time steps. From this, the following conclusion can be drawn: with ample exits, any designated seating for people with disabilities will suffice, so long as the placement is non-random.

\subsection{People Falling}
\begin{figure}[!htbp]
     \includegraphics[width=8.9cm]{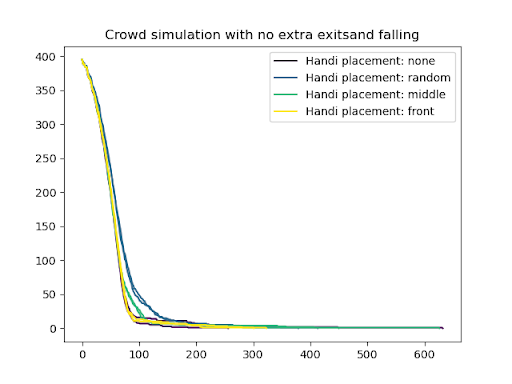}
 \caption{Designated placement of people with disabilities while including crowd members that can fall over in crowd of 400.}
 \label{falling}
 \end{figure}

The final test, includes crowd members that have a probability of 1/1000 of falling down and a second probability of 1/100 of staying down for every frame (ie. time step). All groups in each of their simulations suffered from significantly longer evacuation times due to stragglers - stragglers being crowd members who fell over and stayed down for many frames. Often times in these simulations, only a few crowd members would be left as they stayed down for a long period. This explains the long tails in each group (see Figure \ref{falling}). However, the front placement group, while not optimal, did yield the best results of any group including the control with an average evacuation time of about 300 time steps.
 
\subsection{Restrictive Exits}
\begin{figure}[!htbp]
     \includegraphics[width=8.9cm]{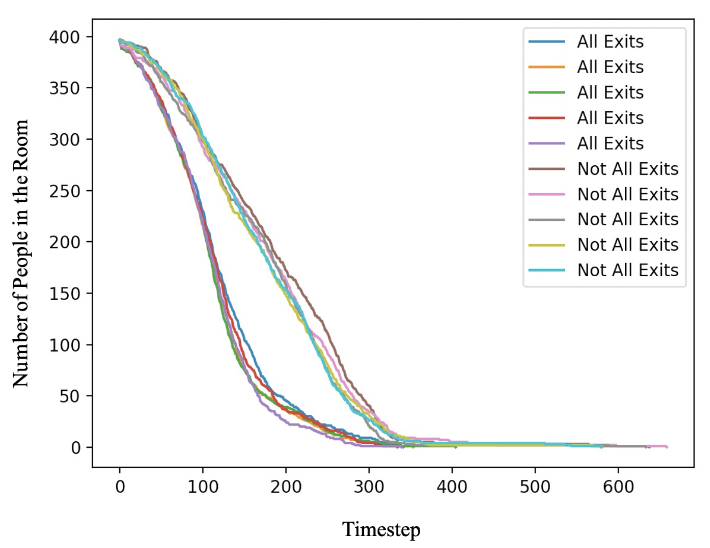}
 \caption{Graph showing the evacuation time when half of the exits are only accessible to people with disabilities.}
 \label{restrictive}
 \end{figure}

Figure \ref{restrictive} shows a pretty non-intuitive result from simulations where we varied who had access to which exits. In this case, we made half of the exits only accessible to agents with disabilities and the other two were only accessible to agents without the 'disability' designation. What we see is that for all five runs where the exits are restricted, the evacuation time was significantly impacted. This shows us that even though it might seem like a good idea to create separate exits, evacuation is most efficient when everyone has access to as many exits as possible. 
\section{Discussion}
Using an agent-based model, we investigated crowd  dynamics and the impact of people with disabilities on the time it takes for the last person to exit the building. First, we looked at the placement of seating for people with disabilities in a grid layout similar to a concert hall with rows of seats. The placement of the specified seats for people with disabilities were either randomly placed throughout the venue, at the front or in the middle of the venue. After running these simulations for each different seating placement, our results showed that when seating for people with disabilities was located at the front of the venue, exit time was the quickest and the most efficient. People with disabilities were able to move towards their nearest exit in the front hand corners of the venue to escape quickly. When the seating placement was random, the able-bodied people were quick to bypass the people with disabilities and they were the last to exit the building causing the exit time to increase. When seating was placed in the middle, the same phenomenon occurred as the random placement. \\
\indent However, when there were an additional two exits placed in the center of the left and right side of the grid, the placement did not seem to significantly affect the exit speed. Although unrealistic, our model had each agent move to their nearest exit and since there were enough exits, everyone was able to leave the building quickly. \\
\indent With the same set up as mentioned above, we included the possibility of agents to fall while in the process of exiting the building. At every time step, a person had a 1/1000 chance they would fall and if they did fall, they had a 1/100 chance they would recover at every time step. The main finding from this addition of people falling is that the location of the fall is more important than the actual fall. In our simulation, if an agent fell in the center of the grid, the other agents were able to quickly move around and get to an exit and the evacuation time was not affected. However, if an agent were to fall right in the middle of the exit, evacuation times are most likely to increase since the agent is obstructing the exit. In future work, it would make sense to change the probability of an agent falling based on their surroundings and how densely packed the area is around them. An agent is more likely to get pushed and fall if they are in the center of a large crowd stampeding to the door. \\
\indent Our next experiment investigated whether the exits should or should not be restricted to certain people. For example, the two exits in the top right and left-hand corners of the grid were only accessible to people without disabilities. In the real world this would simulate if people couldn’t use an exit because there was not exit ramp or other similar stipulations. Our results indicated that restricting any exit for certain people negatively impacts the evacuation time. Making sure every exit is accessible for anyone to leave is a vital component to ensuring evacuation time is as efficient as possible. 

\section{Limitations and Future Work}
Right now, our current ABM is a simplified, toy model of crowd dynamics in emergency situations that has many limitations. The largest limitation of our model is the absence of decision-making architecture for the individual agents. The agents will automatically move to their closest exit which is not realistic in the real world. A person may not know about an exit and go towards one farther away or if an exit is extremely backed up, a person may decide to look for another way out. These are all things that would drastically affect how a crowd exits a building and potentially affect the optimal placement of seating for people with disabilities. Another aspect we would dedicate more time to is a comprehensive parameter search method as well as take a more statistical approach in our analysis. In our current model, we heuristically decided on our parameters but a full search might give us more insightful results. \\
\indent Another limitation with our model is that we did not specify the type of event (concert, seated movie theater, etc.) or the type of emergency taking place. Each event and emergency combination will require different evacuation procedures and we may see different crowd behavior. For example, if the event is a large concert, people will naturally gravitate towards the stage and push people closer together. This will cause a higher crowd density making it harder for people to move in the opposing direction. \\
\indent Some future adaptations for our ABM would include adding decision making architecture as well as investigating chaos and panic in crowd dynamics. How can we model the spread of panic and when does a controlled crowd become a chaotic crowd? These elements could be added to each individual agent to get a better sense of how different crowds evacuate a building.

\nocite{*}

\bibliography{apssamp}

\end{document}